# Influence of *Ln* elements (*Ln* = La, Pr, Nd, Sm) on the structure and oxygen permeability of Ca-containing dual-phase membranes

*Shu Wang[1], Lei Shi[1], Mebrouka Boubeche[1], Xiaopeng Wang[1], Lingyong Zeng[1], Haoqi Wang[1], Zhiang Xie[1], Wen Tan[1], Huixia Luo[1,2,3]\**

[1]*School of Materials Science and Engineering, Sun Yat-Sen University, No. 135, Xingang Xi Road, Guangzhou, 510275, P. R. China*

[2]*Key Laboratory for Polymeric Composite and Functional Materials of Ministry of Education, Sun Yat-Sen University, Guangzhou, 510275, China*

[3]*State Key Laboratory of Optoelecronic Materials and Technologies, Sun Yat-Sen University, Guangzhou, 510275, China*

[\*]*Corresponding author/authors complete details (Telephone; E-mail:)*

*Huixia Luo; (+0086)-2039386124; luohx7@mail.sysu.edu.cn*

**Abstract:** Developing good performance and low-cost oxygen permeable membranes for $CO_2$ capture based on the oxy-fuel concept is greatly desirable but challenging. Despite tremendous efforts in exploring new $CO_2$-stable dual-phase membranes, its presence is however still far from meeting the industrial requirements. Here we report a series of new Ca-containing $CO_2$-resistant oxygen transporting membranes with composition 60wt.%$Ce_{0.9}Ln_{0.1}O_{2-\delta}$-40wt.%$Ln_{0.6}Ca_{0.4}FeO_{3-\delta}$ (C*Ln*O-*Ln*CFO; *Ln* = La, Pr, Nd, Sm) synthesized via a Pechini one-pot method. Our results indicate all investigated compounds are composed of perovskite and fluorite phases, while the perovskite phases in the CNO-NCFO and CSO-SCFO membranes after sintering generates Ca-rich and Ca-less two kinds of grains with different morphologies, where the Ca-less small perovskite grains block the transport of oxygen ions and eventually result in poor oxygen permeability. Among our investigated C*Ln*O-*Ln*CFO membranes, CPO-PCFO exhibits the highest oxygen permeability and excellent $CO_2$ stability, which were mainly associated with the improvement in crystal symmetry, non-negligible electronic conductivity of fluorite phase and the enhancement in electronic conductivity of perovskite. Our results establish Ca-containing oxides as candidate material platforms for membrane engineering devices that combine $CO_2$ capture and oxygen separation.

# 1. Introduction

Mixed oxygen-ion and electronic conducting membranes (MIECMs) are of interest as their potential application in the chemical engineering industry and energy fields [1-4], such as oxygen separation from oxygen-enriched air [5-8], partial oxidation of methane to syngas and water splitting for hydrogen production by selective removing of oxygen [9-11] and so on. In particular, MIECMs have been recently engineered in an effort to realize process intensification [12-15]. At present, one of the famous intensification processes is the oxy-fuel combustion, in which the MIECMs are proposed to be used as the prospect oxygen suppliers and realize $CO_2$ capture at the same time. Generally speaking, oxygen is uninterruptedly transferred from the high oxygen concentration side to the low oxygen concentration side, while part of $CO_2$ as the sweeping gas is recycling, the rest is captured and stored for reducing the emission of $CO_2$ [16,17]. It has been estimated this oxy-fuel route can reduce approximately 60% by energy and 35% by costs in comparison with the conventional cryogenic distillation method [18].

Nonetheless, in the process of practical application, there is a consensus that MIECMs must meet the following main requirements, including good chemical/physical/mechanical stability, high permeability and low-cost and so on. In the search for the MIECM candidates, researchers initially focused on the single-phase ceramic oxides with perovskite structures due to their high oxygen permeabilities. Unfortunately, previous studies have shown that most of single-phase oxygen transport membranes (OTMs) exhibit dissatisfied mechanical/chemical stability, leading to the drop even stop of the oxygen permeation when applied in a severely chemical environment, such as $CO_2/SO_2$ environment [19-21]. For example, oxygen permeation flux through the famous single-phase OTM $Ba_{0.5}Sr_{0.5}Co_{0.8}Fe_{0.2}O_{3-\delta}$ immediately drops to zero when employing $CO_2$ as sweep gas due to carbonates formation [22]. Similar phenomena also have been observed in other single-phase perovskites ($ABO_3$) OTMs, especially in the oxides containing alkaline earth elements in the *A*-site [23]. Thus, researchers draw their attention to the dual-phase OTMs as promising alternatives, which is consisted of a fluorite phase ($CeO_2$ type) as ionic conductor (IC) and a perovskite oxide ($ABO_3$) as electronic conductance phase. As a result, large progress has been made in the research and understanding of dual-phase OTMs. Typical dual-phase membranes with the composition $Ce_{0.8}Gd_{0.2}O_{2-\delta}$-$Pr_{0.6}Sr_{0.4}Co_{0.5}Fe_{0.5}O_{3-\delta}$ (CGO-PSCFO) [24], $Ce_{0.9}Gd_{0.1}O_{2-\delta}$-$Ba_{0.5}Sr_{0.5}Co_{0.8}Fe_{0.2}O_{3-\delta}$ (CGO-BSCFO) [25], $Ce_{0.8}Gd_{0.2}O_{2-\delta}$-$La_{0.7}Sr_{0.3}MnO_3$ (CGO-LSMO) [26] and so on, have also been discovered. Compared with the well-known single-phase perovskite OTMs, most of these aforementioned dual-phase OTMs exhibit much higher $CO_2$-stability but lower oxygen permeability. Thus, it is of still great interest to explore new OTMs with outstanding oxygen permeability and stability.

At present, several significant endeavors have been made to develop Ca-based OTMs. The reason why choosing Ca-based compounds is as follows: (i) there is abundant Ca element on the earth thereby they are low-cost; (ii) the formation of $CaCO_3$ is more difficult than those of $BaCO_3$ and $SrCO_3$ under the similar conditions due to the ionic radii of $Ca^{2+}$ is much smaller than those of $Ba^{2+}$ and $Sr^{2+}$ [27]. Previous reports have documented that Ca-doping single-phase OTMs exhibits higher $CO_2$ stability and comparable oxygen permeability in comparison with the Ba/Sr-containing single-phase OTMs, whereas there are rare Ca-containing dual-phase membranes reported [17,27-31]. Herein, we report a series of calcium-containing composites with the formula 60wt.%$Ce_{0.9}Ln_{0.1}O_{2-\delta}$-40wt.% $Ln_{0.6}Ca_{0.4}FeO_{3-\delta}$ ($Ln$ = La, Pr, Nd, Sm; denoted as CLO-LCFO, CPO-PCFO, CNO-NCFO and CSO-SCFO). The structure, oxygen permeability and stability will be systematically studied in this paper.

## 2. Experimental

### 2.1 Preparation of powder and membranes

60wt.%$Ce_{0.9}Ln_{0.1}O_{2-\delta}$-40wt.%$Ln_{0.6}Ca_{0.4}FeO_{3-\delta}$ (C$Ln$O-$Ln$CFO; $Ln$ = La, Pr, Nd, Sm) were synthesized through a Pechini one-pot method. Firstly, the nitrates with stoichiometric ratios were dissolved in deionized water. Then the citric acids were mixed with the ratio of citric acid to metal cation 2:1 and formed solutions. A few ethylene glycols (~3 mL) as the dispersant then were added to the obtained solution. Gels can be formed after the solution were gradually evaporated and stirred at 110 °C for several hours. Subsequently, the gels were dried in an oven at 140 °C and heated in air at 600 °C for 8 hours to get rid of organic macromolecules. The final dual-phase powders can be obtained after heating in air at 950 °C for 10 hours. Then the as-prepared powders were milled and pressed into pellets in a stainless mold (12 mm i.d.) under uniaxial pressure of ~ 150 MPa. The fresh pellets were sintered at 1400 °C for 5 hours and later were carefully polished to be the thickness of 0.60 mm by using sandpapers (average particle size is 10 μm) to gain the targeted dual-phase membranes.

**2.2 Characterization**

The crystal structures of C*Ln*O-*Ln*CFO powders and dense membranes were analyzed with powder X-ray diffraction (XRD, MiniFlex, Rigaku with Cu Kα) in the 2-theta range 10-120° intervals of 0.02°. In order to determine the space groups and unit cell parameters, Rietveld refinements were performed on the powder diffraction data with the FULLPROF suite software [32]. The surface morphologies of as-prepared membranes were studied by scanning electron microscopy (SEM, Quanta 400F, Oxford and GeminiSEM 500, Carl Zeiss) at the accelerating voltage of 20 kV. Subsequently, the grain compositions of CNO-NCFO and CSO-SCFO membranes were determined by energy dispersive X-ray spectroscopy (EDXS).

**2.3 Oxygen permeability measurement**

The oxygen permeability was tested with a home-made corundum apparatus. The as-prepared membranes were sealed on the alundum conduit using the ceramic glue (Huitian, Hubei, China). The actual area in the container is around 0.709 cm$^2$. Synthetic air (21 vol.% $O_2$ and 79 vol.% $N_2$) with a flow rate of 150 mL min$^{-1}$ flowed into the feeding side of our investigated membranes and 98 vol.% He (99.999 vol.% He)/$CO_2$ (99.999 vol.% $CO_2$) + 2 vol.% Ne (99.999 vol.% Ne, an internal standard gas) with a flow rate of 50 mL min$^{-1}$ flowed into the sweeping side of investigated membranes. All the gas rates were controlled by the mass flow meters (Sevenstar, Beijing, China). The effluent gas compositions were identified by the gas chromatograph (Agilent, 7890B, USA). There is a consensus that it always exits slight oxygen leakage in the test due to the defect of sealing. This part should be subtracted for calculating the amount of oxygen permeability, and the leakage oxygen was no more than 10 % of the total oxygen permeation flux.

**2.4 Calculation of the average metal-oxygen bond energy**

The average metal-oxygen bond energy of perovskite structure ($\langle AB\mathrm{E}\rangle$) was calculated on the basis of the handbook of thermodynamic data [33]. Specifically, the average metal-oxygen bond energy including $A$-O/$B$-O bond energies ($\langle A$-O$\rangle$, $\langle B$-O$\rangle$) within the $AB\mathrm{O}_3$ perovskite structure were calculated according to the following formulas:

$$\langle AB\mathrm{E}\rangle = \langle A\text{-O}\rangle + \langle B\text{-O}\rangle \tag{1}$$

$$\langle A\text{-O}\rangle = x_{A'}\langle A'\text{-O}\rangle + x_{A''}\langle A''\text{-O}\rangle + x_{A'''}\langle A'''\text{-O}\rangle \tag{2}$$

$$\langle B\text{-O}\rangle = x_{B'}\langle B'\text{-O}\rangle + x_{B''}\langle B''\text{-O}\rangle + x_{B'''}\langle B'''\text{-O}\rangle \tag{3}$$

$$\langle A'\text{-O}\rangle = \frac{(\Delta_f H^\circ{}_{A_m O_n} - m\cdot\Delta H s_A - 0.5n\cdot D_{O_2})}{m\times CN_A} \tag{4}$$

$$\langle B'\text{-O}\rangle = \frac{(\Delta_f H^\circ{}_{B_m O_n} - m\cdot\Delta H s_B - 0.5n\cdot D_{O_2})}{m\times CN_B} \tag{5}$$

Where, the $A'$, $A''$, $A'''$ represent the metal atoms occupied in $A$-site of perovskite, whereas the $B'$, $B''$, $B'''$ represent the metal atoms occupied in $B$-site of perovskite. The $x_{A'}$, $x_{A''}$, $x_{A'''}$ and $x_{B'}$, $x_{B''}$, $x_{B'''}$ denote the molar fractions of corresponding metal atoms in the $A$-site and $B$-site of perovskite, respectively. $CN_A$ and $CN_B$ represent the coordination numbers of *A*-site and *B*-site, respectively. $\Delta_fH°A_mO_n$, $\Delta_fH°B_mO_n$ denote the enthalpies of formation of $A(B)mOn$ at 298.15 K, respectively. And $\Delta Hs_A$ and $\Delta Hs_A$ are the enthalpies of sublimation of corresponding elemental metal at 298.15 K, respectively. $D_{O2}$ is dissociation energy (500.2 kJ/mol).

## 3. Result and discussion

### 3.1 Phase structure of powder

**Figs. 1a-d** show the XRD patterns together with the Rietveld fitting profiles for the as-obtained C*Ln*O-*Ln*CFO (*Ln* = La, Pr, Nd, Sm) composite powders. The data reveal that all our investigated dual-phase powders after calcined in 950 °C for 10 hours only consist of the cubic fluorite phase (space group: $Fm\bar{3}m$, *No.* 255) and the pseudo-cubic ($a \times b \times c \approx \sqrt{2}a_c \times \sqrt{2}a_c \times 2a_c$, where *a, b, c,* represent lattice constants of orthorhombic structure, and $a_c$ represent the lattice constant of cubic structure [34]) perovskite phase (space group *Pbnm*, *No.* 62). The lattice constants for our investigated compounds are listed in **Table 1.** For comparison, the converted lattice parameters and volume are further plotted in **Figs. 2a and b**. It can be seen that the cell volumes in perovskite decrease with the increases of the atomic numbers, which can be attributed to the decrease of the corresponding ion radii (see **Table 2**, extracted from the Shannon's and Jia's table [34, 35]); while the lattice parameter value for the fluorite phase in the CPO-PCFO is the smallest among these investigated compounds. It can be explained that Pr elements have multivalent such as +3 and +4, and $Pr^{4+}$ (96 pm) has the smallest ionic radius in the eight coordination, whereas other rare-earth elements

(La, Sm, Nd) only have a fixed valence (+3) in the eight coordinations. This change trend is in agreement with that in the single-phase fluorite phase $Ce_{1-x}Ln_xO_{2-\delta}$ system ($Ln$ = Gd, La, Tb, Pr, Eu, Er, Yb, Nd) reported by Balaguer *et al.* [36].

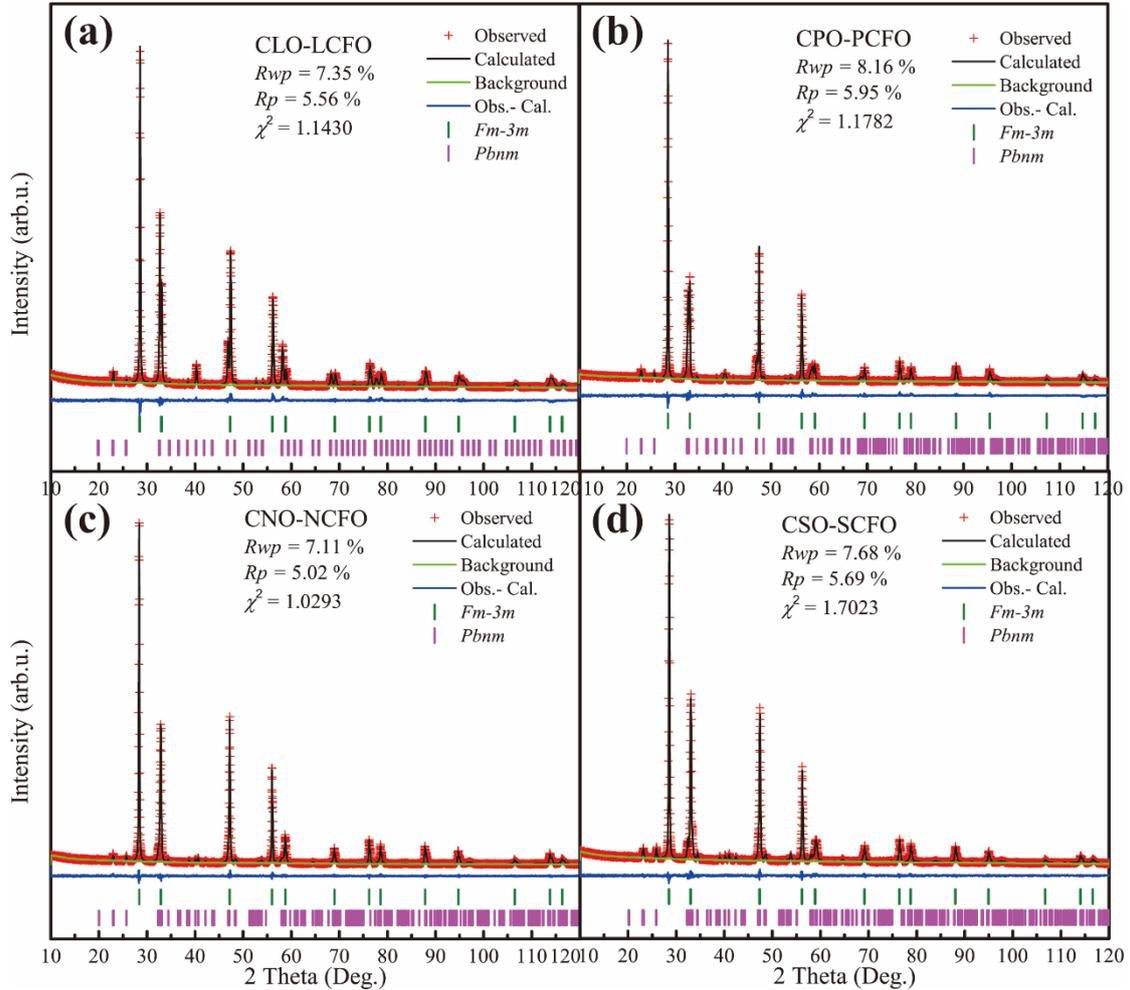

**Fig. 1** XRD patterns with Rietveld refinements for (a) 60wt.%$Ce_{0.9}La_{0.1}O_{2-\delta}$-40wt.%$La_{0.6}Ca_{0.4}FeO_{3-\delta}$, (b) 60wt.%$Ce_{0.9}Pr_{0.1}O_{2-\delta}$-40wt.%$Pr_{0.6}Ca_{0.4}FeO_{3-\delta}$, (c)60wt.%$Ce_{0.9}Nd_{0.1}O_{2-\delta}$-40wt.%$Nd_{0.6}Ca_{0.4}FeO_{3-\delta}$, and (d) 60wt.%$Ce_{0.9}Sm_{0.1}O_{2-\delta}$-40wt.%$Sm_{0.6}Ca_{0.4}FeO_{3-\delta}$ powders.

In order to observe the evolution of the cell parameters for perovskite phases intuitively, we divide *a*, *b* by √2 and *c* by 2 (Based on the pseudo-cubic structure, $a \times b \times c \approx \sqrt{2}a_c \times \sqrt{2}a_c \times 2a_c$). As shown in **Fig. 2a**, it can be found that the difference of converted cell parameters (a/√2, b/√2, c/2) in perovskite phases increases with the increase of atomic numbers. Specifically, the a/√2, b/√2 and c/2 are almost the same in the case of CLO-LCFO, whereas it is very different in the CSO-SCFO compound. That is to say, the symmetries of perovskite in composite powders are getting worse as the increase of atomic numbers of rare earth elements. Also, the tilting angle Φ (refer to the previous literature for the specific calculation process) is an important geometrical parameter to evaluate the distortion extent in the orthorhombic perovskite structure [37]. As the Φ value gets closer to 0, the distorted Fe-O octahedron will be more closer to the ideal normal octahedron. In other words, the crystal structure is more likely to change to a stable cubic structure. The calculated Φ values for our investigated compounds are summarized in **Table 1**. Our results show that the Φ gradually increases from 3.41 ° to 15.49 ° with the atom number increasing, indicating that the crystal symmetry becomes worse as the atom numbers are getting bigger.

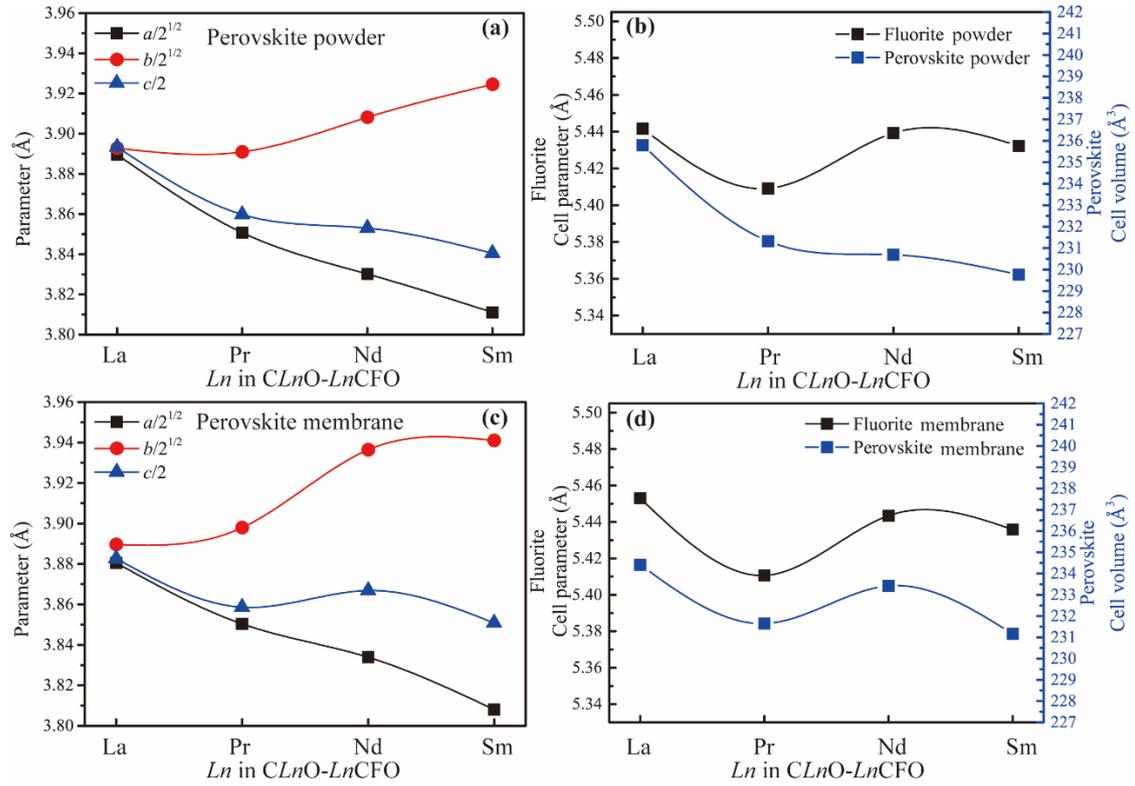

**Fig. 2** The evolution of the cell parameters for C*Ln*O-*Ln*CFO **(a)(b)** powders and **(c)(d)** membranes.

**Table 1** Refinement results from XRD patterns for 60wt.%$Ce_{0.9}Ln_{0.1}O_{2-\delta}$-40wt.%$Ln_{0.6}Ca_{0.4}FeO_{3-\delta}$ (Ln = La, Pr, Nd, Sm) powders after calcined at 950 °C for 5 hours.

| composition | PCFO Cell Parameter | | | | CPO Cell Parameter | $\chi^2$ | $R_{wp}$ / % | $R_p$ / % |
| --- | --- | --- | --- | --- | --- | --- | --- | --- |
| | $a$/Å | $b$/Å | $c$/Å | $\Phi$/° | $a$/Å =$b$/Å =$c$/Å | / % | | |
| CLO-LCFO | 5.5007(17) | 5.5052(11) | 7.7866(17) | 3.41 | 5.4417(4) | 1.1430 | 7.35 | 5.56 |
| CPO-PCFO | 5.4458(9) | 5.5026(7) | 7.7195(12) | 9.11 | 5.40912(19) | 1.1782 | 8.16 | 5.95 |
| CNO-NCFO | 5.4166(7) | 5.5269(7) | 7.7060(11) | 13.04 | 5.43923(15) | 1.0293 | 7.11 | 5.02 |
| CSO-SCFO | 5.3897(7) | 5.5501(8) | 7.6809(12) | 15.49 | 5.4322(4) | 1.7023 | 7.68 | 5.69 |

**Table 2** Effective ionic radii for some elements according to the coordination number.

| Phase | cation | coordination number | effective ionic radii (pm) |
| --- | --- | --- | --- |
| Fluorite | $Ce^{4+}$ | 8 | 97 |
| | $Ce^{3+}$ | 8 | 114.3 |
| | $La^{3+}$ | 8 | 116.0 |
| | $Pr^{3+}$ | 8 | 112.6 |
| | $Pr^{4+}$ | 8 | 96 |
| | $Nd^{3+}$ | 8 | 110.9 |
| | $Sm^{3+}$ | 8 | 107.9 |
| Perovskite | $La^{3+}$ | 12 | 136.0 |
| | $Pr^{3+}$ | 12 | 132 |
| | $Nd^{3+}$ | 12 | 127 |
| | $Sm^{3+}$ | 12 | 124 |

Now we turn to explore the phase stability of investigated composite powders under pure Ar or $CO_2$ environment. **Figs. 3a and b** depict the partial enlarged XRD patterns of four investigated powders after heating at 950 °C for 20 hours under pure Ar and $CO_2$, where the black lines are the partial enlarged XRD patterns of the corresponding fresh composite powders. The insert images are the normal scale XRD patterns of our investigated powders. Compared with the fresh powders, the treated powders keep the same crystal structures as before. It is obvious that no additional diffraction peaks arise and all reflection peaks are assigned to the perovskite and fluorite phase no matter in Ar or He atmosphere, which implies our investigated dual-phase powders are stable under both low oxygen partial pressure and pure $CO_2$ environments. In addition, **Fig. 3c** show the partial enlarged XRD patterns after heating at 800 °C for 1 hour under reducing atmosphere (5 vol.% $H_2$ + 95 vol.%Ar). Comparing with the fresh powders (black lines), there is no obvious impurity observed and dual-phase powders only consist of the cubic fluorite phase and perovskite phase. It also indicates that our dual-phase powders are stable under reducing atmosphere.

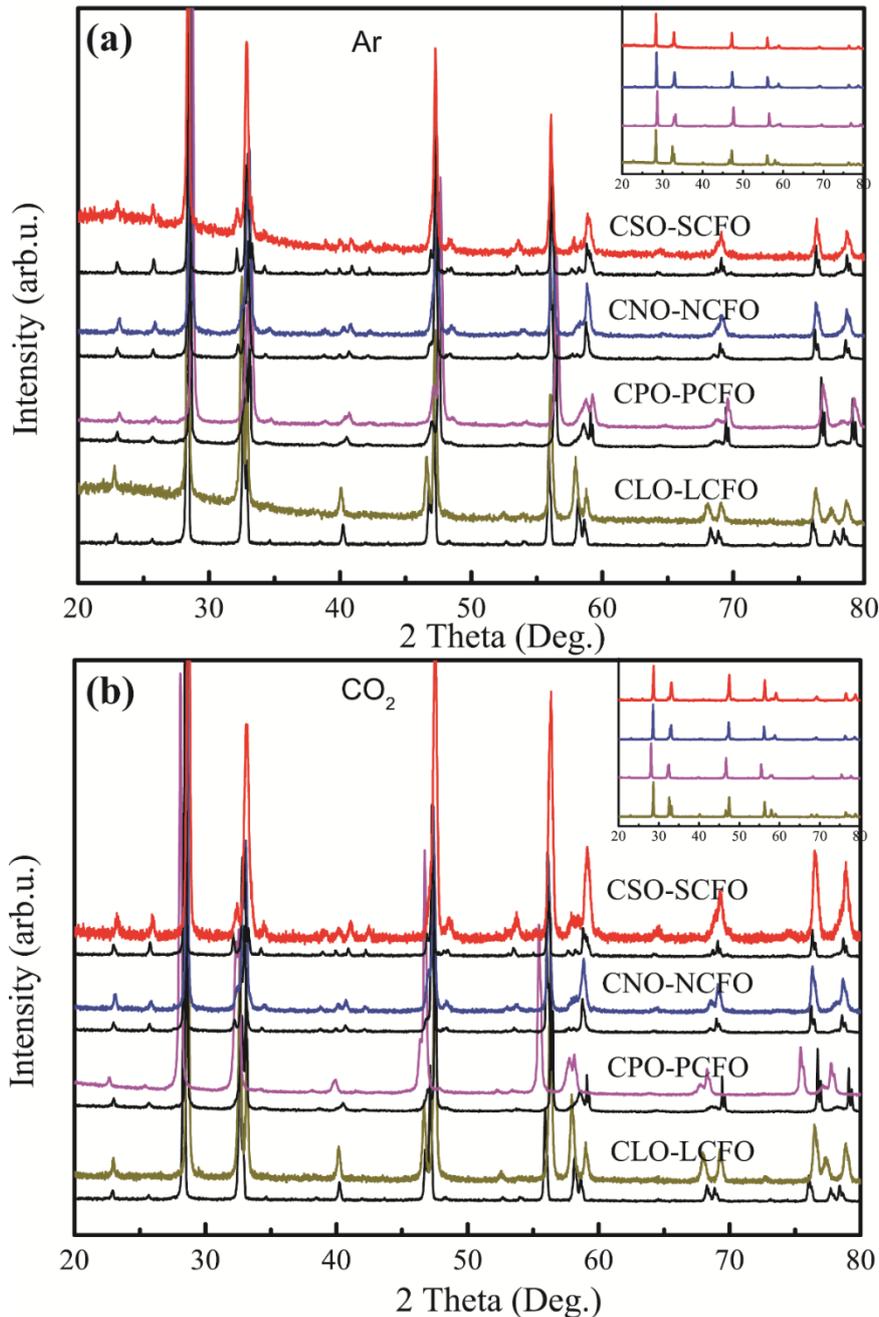

**Fig. 3** XRD patterns of 60wt.%Ce$_{0.9}$Ln$_{0.1}$O$_{2-\delta}$-40wt.%Ln$_{0.6}$Ca$_{0.4}$FeO$_{3-\delta}$ (Ln = La, Pr, Nd, Sm) dual-phase powders after exposed to **(a)** pure Ar and **(b)** pure CO$_2$ at 950 °C for 20 hours (the black line is the partial enlarged XRD pattern of the corresponding initial dual-phase powder. The insert image is the normal scale XRD Pattern of four kind powders.)

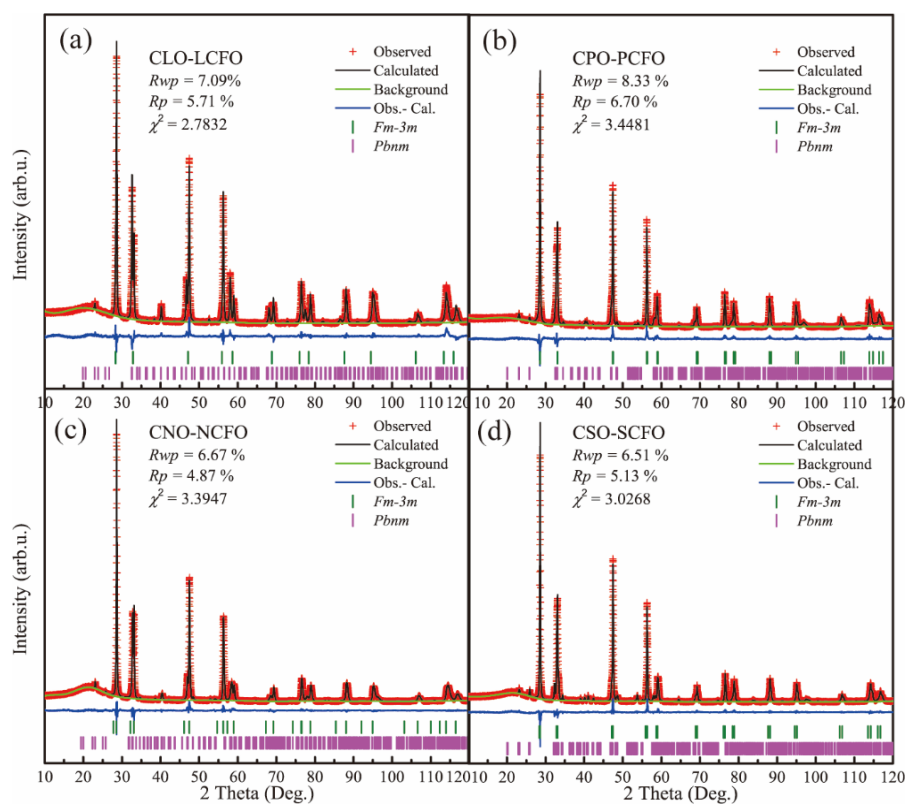

**Fig. 4** XRD patterns with Rietveld refinement of 60wt.%$Ce_{0.9}Ln_{0.1}O_{2-\delta}$-40wt.%$Ln_{0.6}Ca_{0.4}FeO_{3-\delta}$ (*Ln* = La, Pr, Nd, Sm) dual-phase powders after exposed to pure $CO_2$ at 800 °C for 24 hours.

Due to the low decomposition temperature of calcium carbonate (about 850 °C in 1 bar $CO_2$ atmosphere), the stability of dual-phase powders under pure $CO_2$ environment at the lower temperature (800 °C) is further tested. As shown in **Fig. 4**, it can be found that all powders only consist of the cubic fluorite phase (space group: $Fm\bar{3}m$, *No.* 255) and the orthorhombic perovskite phase (space group *Pbnm*, *No.* 62). Also, there is no obvious carbonate formation, which further confirms tour investigated dual-phase powders are $CO_2$ stable at the lower temperature.

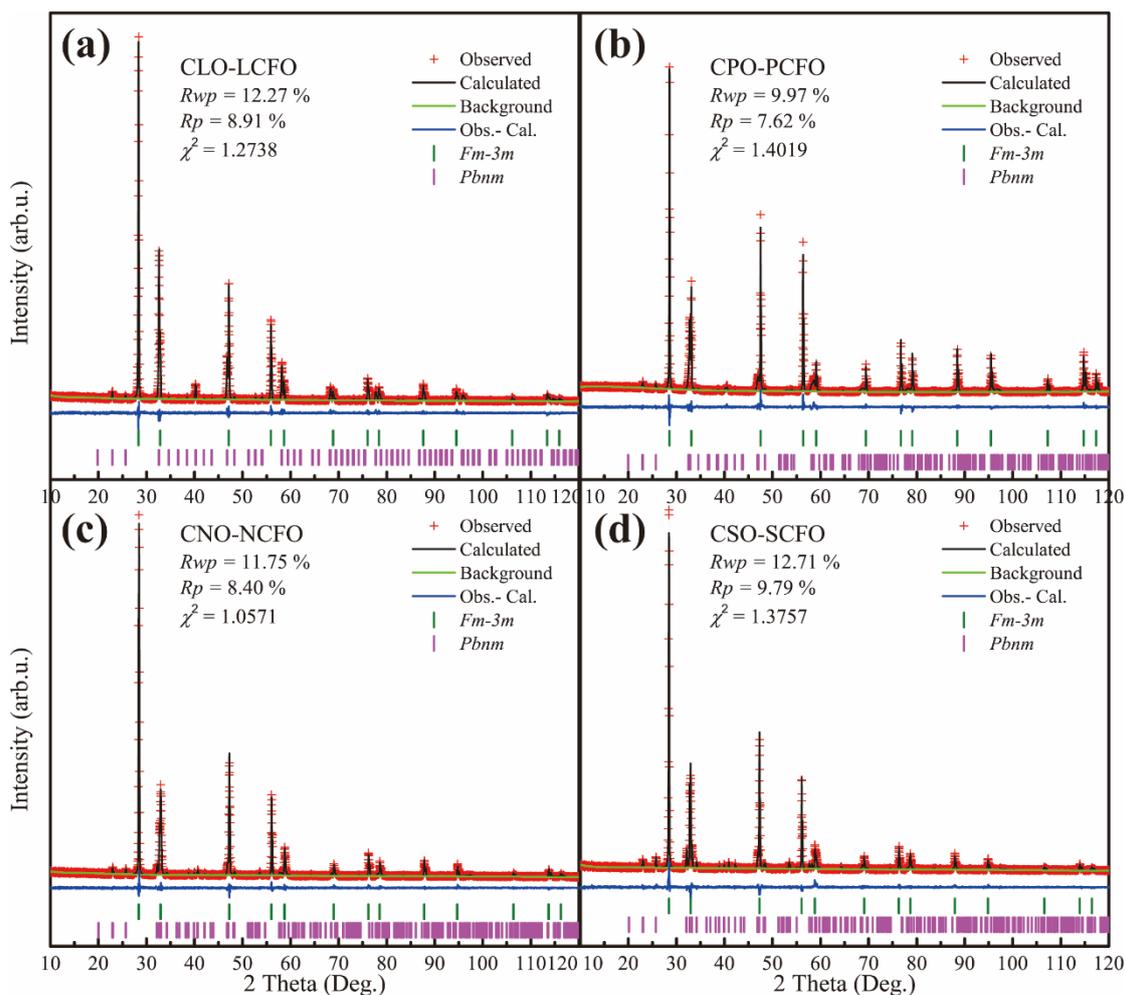

**Fig. 5** XRD patterns with Rietveld refinement of 60wt.%$Ce_{0.9}Ln_{0.1}O_{2-\delta}$-40wt.%$Ln_{0.6}Ca_{0.4}FeO_{3-\delta}$ (*Ln* = La, Pr, Nd, Sm) membranes after sintering at 1400 °C.

## 3.2 Phase structure of membrane

To get the overview of the structures of our investigated membranes, XRD is used to characterize the membranes after sintering at 1400 ºC for 5 hours. As is illustrated in

**Figs. 5a-d**, all diffraction peaks can be assigned into the perovskites phases or fluorite phases without any additional diffraction peaks (such as brownmillerite) [38]. According to our refinement results, the same space groups of perovskite (space group *Pbnm*, *No.* 62) and fluorite (space group: $Fm\overline{3}m$, *No.* 255) phases for all membranes are maintained after high-temperature sintering.

**Table 3** Refinement results from XRD patterns for 60wt.%$Ce_{0.9}Ln_{0.1}O_{2-\delta}$-40wt.%$Ln_{0.6}Ca_{0.4}FeO_{3-\delta}$ (*Ln* = La, Pr, Nd, Sm) dual-phase membranes after sintering at 1400 °C for 5 hours.

| composition | PCFO Cell Parameter | | | | CPO Cell Parameter | $\chi^2$ | $R_{wp}$ / % | $R_p$ / % |
|---|---|---|---|---|---|---|---|---|
| | $a$/Å | $b$/Å | $c$/Å | $\Phi$/° | $a$/Å =$b$/Å =$c$/Å | | | |
| CLO-LCFO | 5.4879(5) | 5.5007(5) | 7.7654(8) | 4.35 | 5.45306(18) | 1.2738 | 12.27 | 8.91 |
| CPO-PCFO | 5.4452(10) | 5.5125(8) | 7.7173(14) | 9.71 | 5.41066(8) | 1.4019 | 9.97 | 7.62 |
| CNO-NCFO | 5.4219(6) | 5.5669(7) | 7.7336(11) | 15.06 | 5.44343(18) | 1.0571 | 11.75 | 8.40 |
| CSO-SCFO | 5.3854(4) | 5.5734(4) | 7.7016(6) | 17.15 | 5.4358(2) | 1.3757 | 12.71 | 9.79 |

**Table 3** shows the refinement results of dual-phase membranes after sintering at 1400 °C for 5 hours. The effect of rare earth elements on the perovskite structure for the composite membranes is good agreement with the dual-phase powders, that the crystal symmetry become worse with the atom number increasing. Compared with the powder samples, the tilting angles ($\Phi$s) for all membranes are gettting larger, indicating all structure symmetries for the perovskite phase in the membranes become worse after sintering at 1400 °C for 5 hours. In this sense, the octahedral structure becomes more deformed. These changes are also reflected in the XRD patterns, **Figs. 6a-h** shows the close-up of the XRD patterns at 31 ° ≤ 2Θ ≤ 34 °. As for the pseudo-cubic perovskite (space group *Pbnm*, *No.* 62,), the diffraction angle of (200), (020) and (112) lattice plane should be the same for the ideal cubic structure. In other words, the closer the diffraction angle of (200), (020) and (112) lattice plane is, the closer the crystal structure is to the ideal cubic structure. By observing the change of diffraction angle of crystal plane, it can be found that the symmetry for CLO-LCFO and CPO-PCFO samples is better than that of CNO-NCFO and CSO-SCFO no matter in powder or membrane samples. Moreover, all membranes are better than these of powders. This changing trend is good accordance with the result of the tilting angle Φ. As mentioned above, **Fig. 2c and d** show the evolution of cell parameters vs lanthanide elements for the dual-phase membranes. It is worth noting that the perovskite phase in the Pr-containing dual-phase membrane also exhibits a similar mutation as the fluorite phase, which is likely to be caused by the partial conversion of $Pr^{3+}$ to $Pr^{4+}$.

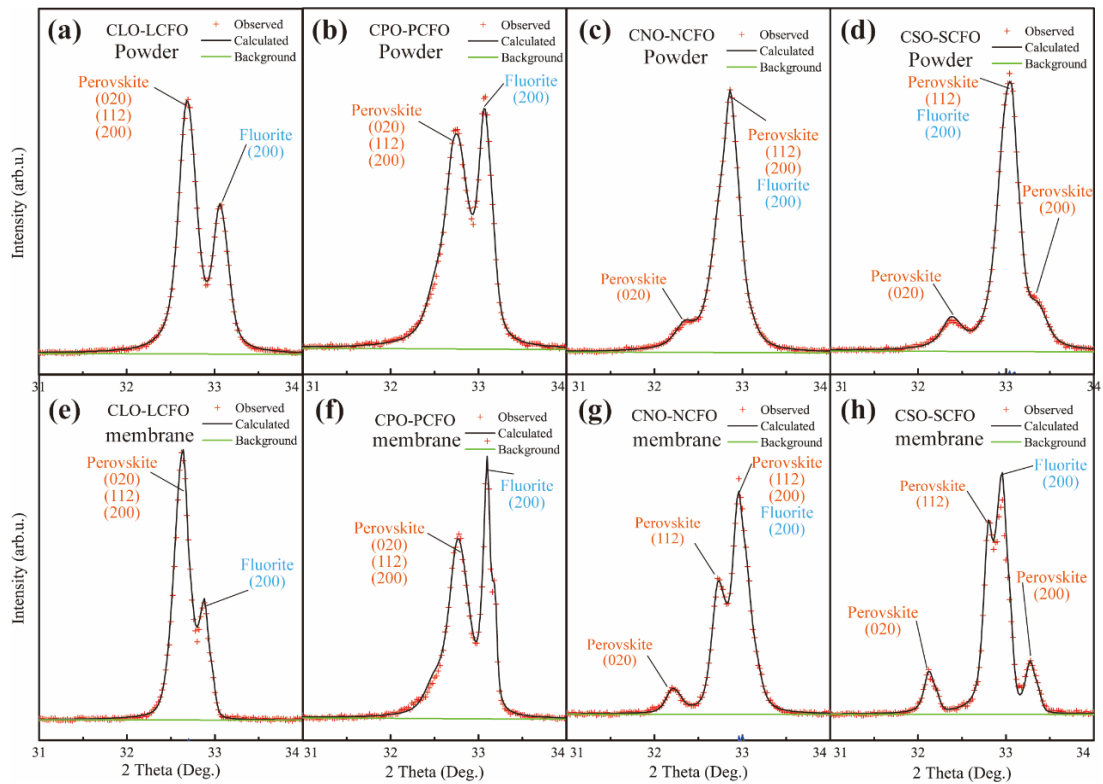

**Fig. 6** The close-up of XRD patterns of 60wt.%$Ce_{0.9}Ln_{0.1}O_{2-\delta}$-40wt.%$Ln_{0.6}Ca_{0.4}FeO_{3-\delta}$ (Ln = La, Pr, Nd, Sm) powder and membranes at 31 ° ≤ 2Θ ≤ 34 °.

**3.3 Membrane morphology**

It has been generally accepted that foreign impurity or formation of the third phases will block the IC/EC continuous paths, resulting in hindering the oxygen separation. In order to further examine the chemical compatibility of our membranes, the scanning electron microscope (SEM) has been conducted. **Fig. S1** exhibits the secondary electron (SE, **Fig. S1a, c, e, g**) and backscattered electron (BSE, **Fig. S1b, d, f, h**) images of C*Ln*O-*Ln*CFO (*Ln* = La, Pr, Nd, Sm) composite membranes after high-temperature sintering. As is shown in **Fig. S1**, all membranes are densely packed and no cracks were observed, only a small number of blind holes are detected. However, the grain sizes are different from each other. In terms of CLO-LCFO and CPO-PCFO membranes, the grains sizes for two phases are very similar (~ 1.5 μm), and the grains are distributed uniformly. Nevertheless, in addition to the crystalline grains with the above size, the grains with the smaller size (~ 0.4 μm) also appear in CNO-NCFO and CPO-PCFO membranes.

To gain more insight into compositions and element distributions in the C*Ln*O-*Ln*CFO membranes studied here, the energy dispersive X-ray spectrometer (EDXS) was conducted. **Fig. 7** shows the SE and EDXS mapping overlap images of dual-phase membranes after sintering at 1400 °C for 5 hours in air. The specific distribution mappings of each element are shown in **Fig. S2-S5**. It is clear in **Fig. 8b, d, f, h** that the CLO-LCFO and CPO-PCFO are comprised of fluorite and perovskite two phases, however, the third phase occurs in the CSO-SCFO and CNO-NCFO membranes. In order to further explore the composition of each phase, the EDXS point analysis is performed in our investigated membranes (no less than 5 sampling points for each phase). The converted data from EDXS point analysis are summarized in **Table 4**. It may be noted that a small amount of calcium diffuses into the fluorite phase, whereas iron exists only in the perovskite phase for these four membranes. In the case of the perovskite phases in CLO-LCFO and CPO-PCFO membranes, both chemical compositions contain cerium element (ii: $La_{0.534(4)}Ca_{0.33(2)}Ce_{0.065(4)}Fe_{1.06(2)}O_{3-\delta}$, iv: $Pr_{0.473(6)}Ca_{0.36(1)}Ce_{0.088(6)}Fe_{1.08(2)}O_{3-\delta}$). However, there are two components in the perovskite phases of CNO-NCFO and CSO-SCFO membranes. One perovskite grains with the larger size, (label: vi and ix) are the Ca-rich phases (vi: $Nd_{0.394(3)}Ca_{0.45(3)}Ce_{0.08(1)}Fe_{1.06(4)}O_{3-\delta}$, iv: $Sm_{0.27(2)}Ca_{0.62(4)}Ce_{0.05(1)}Fe_{1.05(1)}O_{3-\delta}$), while the smaller ones (label: vii and x) are the Ca-less phases (vii: $Nd_{0.54(3)}Ca_{0.22(1)}Ce_{0.160(8)}Fe_{1.08(3)}O_{3-\delta}$, x: $Sm_{0.51(2)}Ca_{0.25(1)}Ce_{0.19(2)}Fe_{1.04(4)}O_{3-\delta}$). Previous studies showed that more calcium doping can promote the growth of grain size during high-temperature sintering. For example, the grain size is obviously getting bigger when the Ca content are higher in $La_{1-x}Ca_xFeO_3$ (e.g. $La_{0.9}Ca_{0.1}FeO_{3-\delta}$ (1.8 μm), $La_{0.8}Ca_{0.2}FeO_{3-\delta}$ (2.2 μm) and $La_{0.7}Ca_{0.3}FeO_{3-\delta}$ (3.8 μm)) ceramics sintered at 1320 °C [39]. Similarly, in our compound, it has been observed that the wave stripe (marked as "*") of near-melting crystalline grain (~ 2 μm) in the Ca-rich perovskite phase, while the grain size is generally smaller (~ 0.4 μm) for the Ca-less perovskite phase.

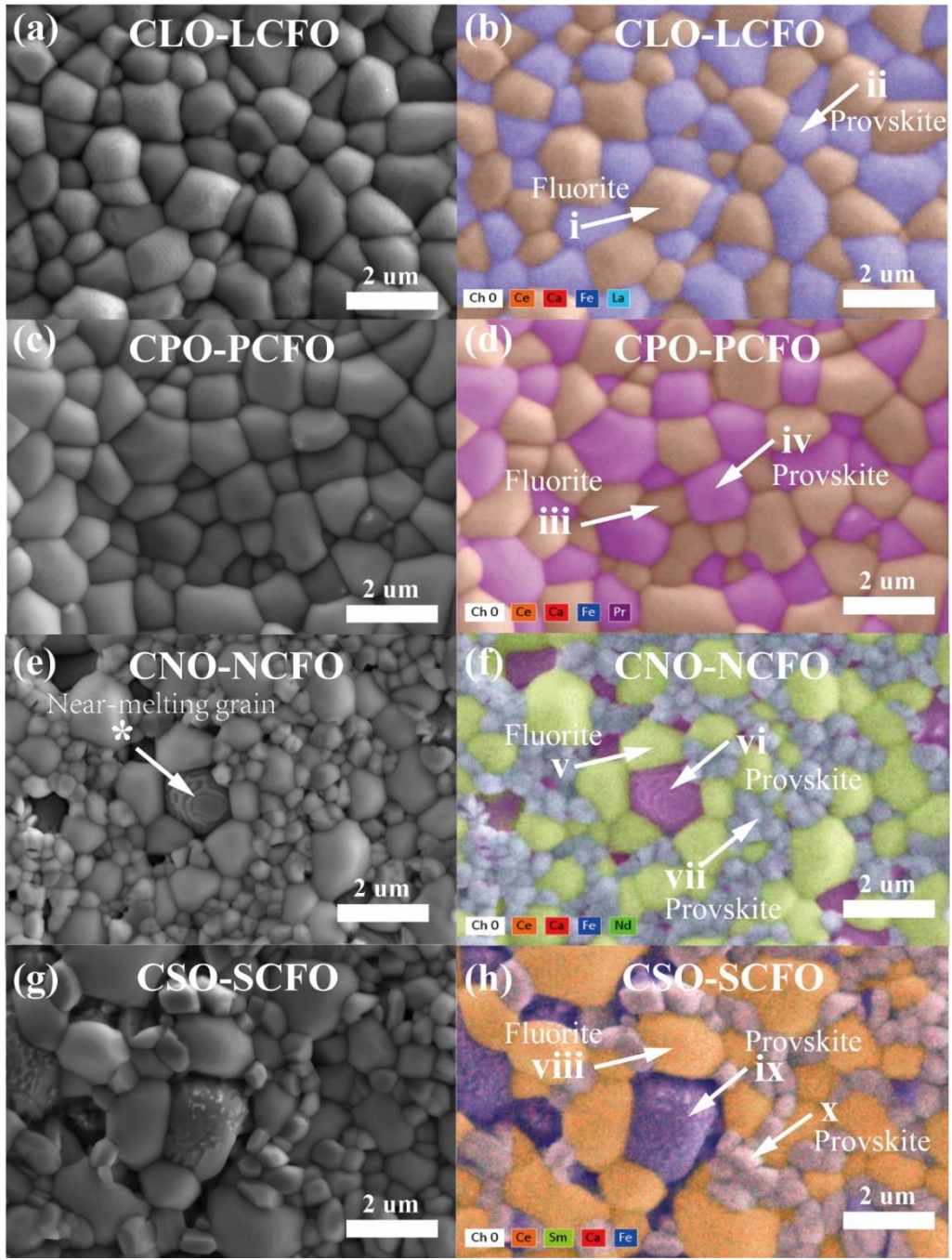

**Fig. 7** SE (a, c, e, g) and EDXS mapping overlap images (b, d, f, h) of fresh 60wt.% $Ce_{0.9}Ln_{0.1}O_{2-\delta}$-40wt.%$Ln_{0.6}Ca_{0.4}FeO_{3-\delta}$ ($Ln$ = La, Pr, Nd, Sm) composites. (Color online only)

**Table 4** EDXS analysis results of $Ce_{0.9}Ln_{0.1}O_{2-\delta}$-40wt.%$Ln_{0.6}Ca_{0.4}FeO_{3-\delta}$ membranes.

| Membrane | Fluorite | | Perovskite | |
|---|---|---|---|---|
| | symbol | composition | symbol | composition |
| CLO-LCFO | i | $Ce_{0.818(5)}La_{0.128(5)}Ca_{0.053(3)}O_{2-\delta}$ | ii | $La_{0.534(4)}Ca_{0.33(2)}Ce_{0.065(4)}Fe_{1.06(2)}O_{3-\delta}$ |
| CPO-PCFO | iii | $Ce_{0.81(1)}Pr_{0.160(5)}Ca_{0.022(6)}O_{2-\delta}$ | iv | $Pr_{0.473(6)}Ca_{0.36(1)}Ce_{0.088(6)}Fe_{1.08(2)}O_{3-\delta}$ |
| CNO-NCFO | v | $Ce_{0.81(2)}Nd_{0.157(3)}Ca_{0.024(4)}O_{2-\delta}$ | vi | $Nd_{0.394(3)}Ca_{0.45(3)}Ce_{0.08(1)}Fe_{1.06(4)}O_{3-\delta}$ |
| | | | vii | $Nd_{0.54(3)}Ca_{0.22(1)}Ce_{0.160(8)}Fe_{1.08(3)}O_{3-\delta}$ |
| CSO-SCFO | viii | $Ce_{0.79(2)}Sm_{0.18(1)}Ca_{0.023(9)}O_{2-\delta}$ | ix | $Sm_{0.27(2)}Ca_{0.62(4)}Ce_{0.05(1)}Fe_{1.05(1)}O_{3-\delta}$ |
| | | | x | $Sm_{0.51(2)}Ca_{0.25(1)}Ce_{0.19(2)}Fe_{1.04(4)}O_{3-\delta}$ |

**3.4 Oxygen permeation performance**

We now focus on the study of the oxygen permeability through our investigated C*Ln*O-*Ln*CFO (*Ln* = La, Pr, Nd, Sm) dual-phase membranes at different temperatures under air/He gradient. For comparison, all our investigated membranes are polished into the same thickness (0.6 mm). Besides, here the data of CPO-PCFO is extracted from the previous report [40]. As is shown in **Fig. 8a**, the oxygen permeation flux enhances with the measuring temperature increasing, which is attributed to the acceleration of oxygen bulk diffusion and strengthen of surficial exchange reactions as the temperature increases [41]. For better comparison, we subsequently plotted the oxygen permeation flux vs *Ln* component in **Fig. 8b**. Clearly, the CNO-NCFO and CSO-SCFO membranes possess lower oxygen permeation fluxes than those of CPO-PCFO and CLO-LCFO membranes. It may be explained as follows: **(a)** the hinder effect of small grain: When two phases are distributed uniformly and form percolation network, which is good for the transport of electron and oxygen ion through membranes. In other words, oxygen permeation is facilitated when the oxides respectively form a continuous network [38]. The oxygen ionic conductor path is continuous in the CLO-LCFO and CPO-PCFO membranes, only a few isolated "islands" are observed, thereby exhibiting better oxygen permeability. However, many ionic grains become isolated "islands" (**Figs. 9c and d**) in the CSO-SCFO and CNO-NCFO membranes owing to the formation of a third phase, resulting in the fluorite phase channel blockage and thereby exhibiting lower oxygen permeability. Besides, the crystal symmetry affects the oxygen permeability. At present, it is widely accepted that the high-symmetry cubic phase is beneficial for the oxygen permeation, since the oxygen ion can be delivered in 3-dimension in the cubic phase, whereas restricted direction in the orthorhombic phase.

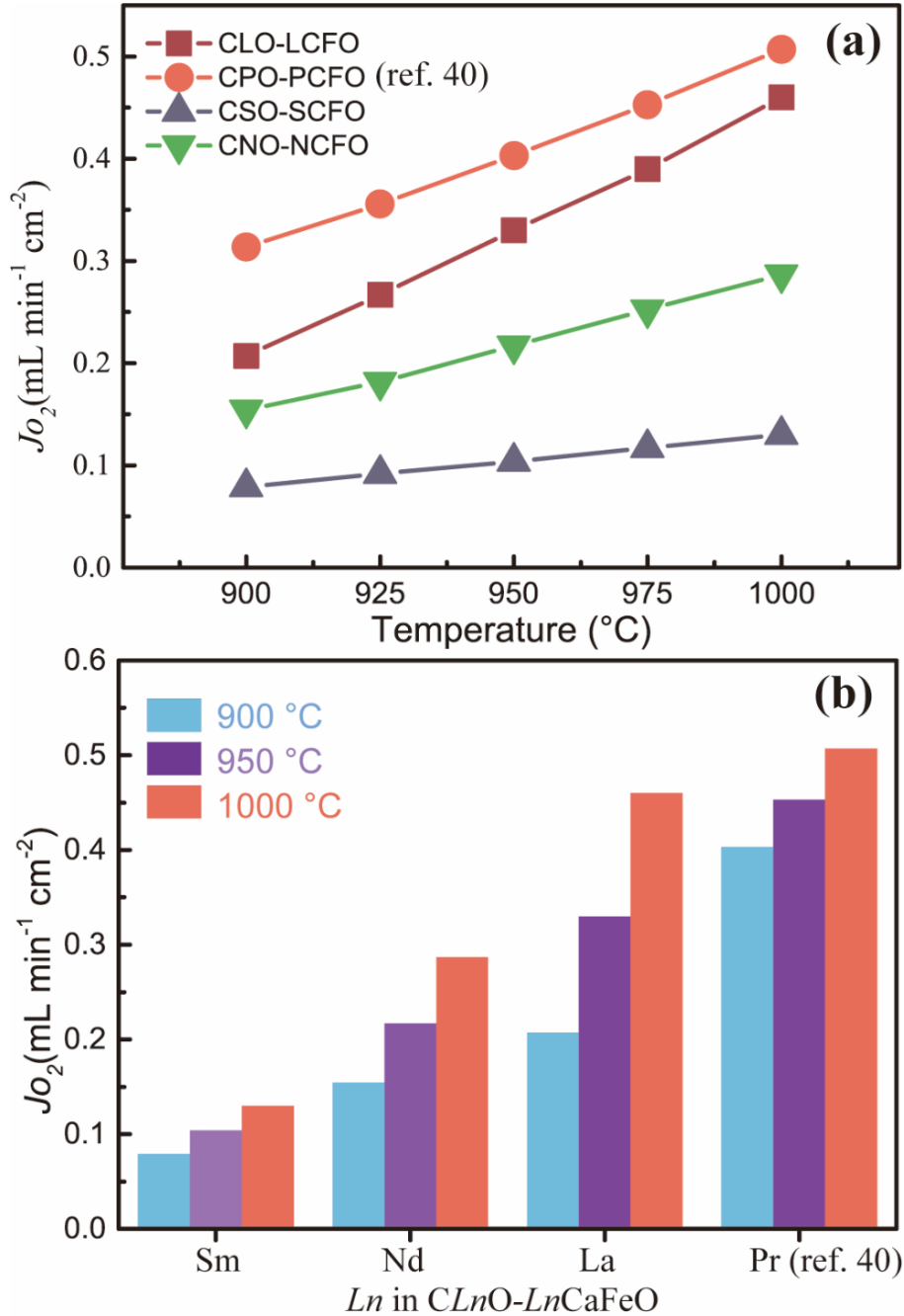

**Fig. 8** Oxygen permeation fluxes through 60wt.%$Ce_{0.9}Ln_{0.1}O_{2-\delta}$-40wt.%$Ln_{0.6}Ca_{0.4}FeO_{3-\delta}$ ($Ln$ = La, Pr, Nd, Sm) dual-phase membranes with pure He. *Conditions: membrane thickness: 0.6 mm; 150 mL $min^{-1}$ air as the feed gas; 49 mL $min^{-1}$ He as sweep gas; 1 mL $min^{-1}$ Ne for calibration.*

For example, oxygen ion can easily migrate on *ab* plane while difficultly along C axis in orthorhombic $La_{0.64}(Ti_{0.92}Nb_{0.08})O_{2.99}$ compound [42]. As a result, CPO-PCFO and CLO-LCFO membranes with higher space symmetry exhibit higher oxygen permeability than those of CNO-NCFO and CSO-SCFO with lower space symmetry on the basis of our XRD refinement oxygen permeation results. This is also consistent with the above analysis.

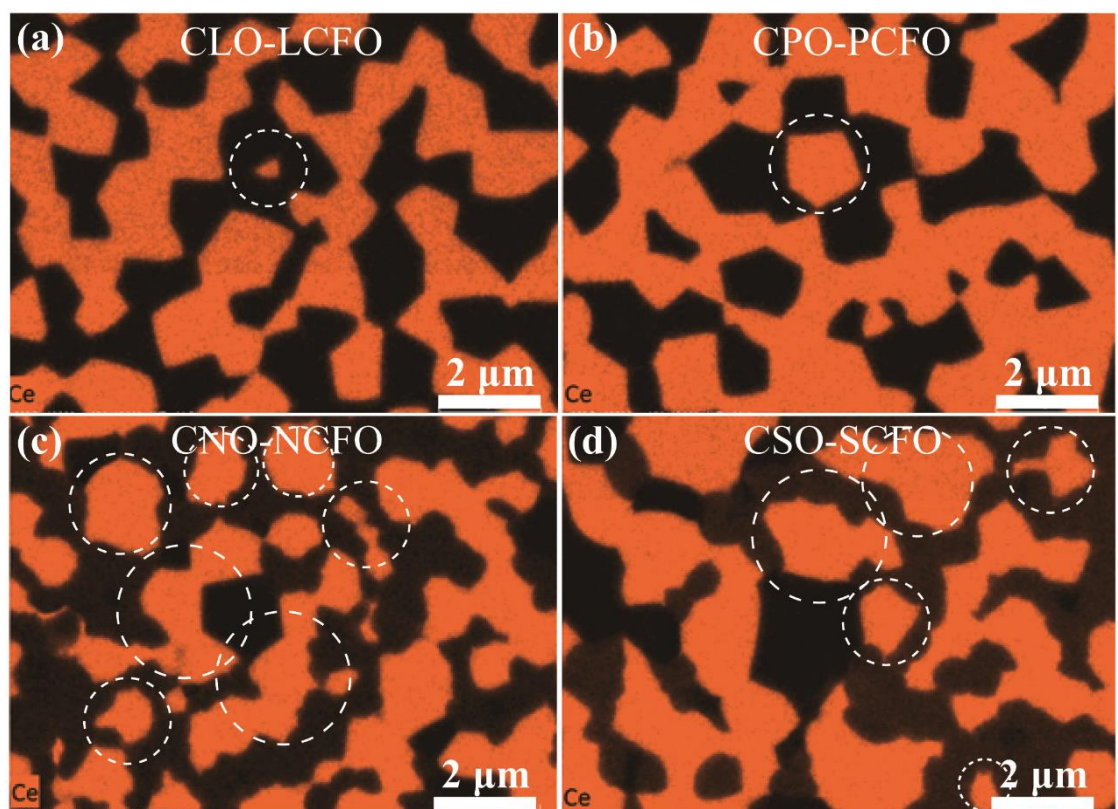

**Fig. 9** Cerium distribution map of $Ce_{0.9}Ln_{0.1}O_{2-\delta}$-40wt.%$Ln_{0.6}Ca_{0.4}FeO_{3-\delta}$ (*Ln* = La, Pr, Nd, Sm) membranes. (Color online only)

It is noticeably seen that the CPO-PCFO composite exhibits the highest oxygen permeability among our investigated four C*Ln*O-*Ln*CFO membranes and yields an oxygen permeation flux of 0.51 mL cm$^{-2}$ min$^{-1}$ at 1000 °C. On the one hand, it may be in connection with the non-negligible electronic conductivity of CPO, whereas other lanthanide elements in C*Ln*O only own the fixed valence (+3). As the Pr element has mixed valence of $Pr^{3+}$ and $Pr^{4+}$, some polarons can hop between different valence states, resulting in improvement in the electronic conductivity of $Ce_{0.9}Pr_{0.1}O_{2-\delta}$ phase [43,44]. For example, the electronic conductivity of $Ce_{0.8}Pr_{0.2}O_{2-\delta}$ is the 0.021 S/cm at 800 °C in the oxygen partial pressure gradient 1/0.21 bar, which is the same order of magnitude comparing with the ionic conductivity (0.077 S/cm) [45]; whereas other electronic conductivity of lanthanide-doped ceria is negligible, such as $Ce_{0.9}Gd_{0.1}O_{2-\delta}$ phase ($\sim 10^{-3.8}$ S/cm) at 800 °C [46]. Since $Ce_{0.9}Pr_{0.1}O_{2-\delta}$ owns both high ionic conductor and non-negligible electronic conductivity, the reaction sites for oxygen permeation cover the whole membrane surfaces for CPO-PCFO, which is not confined to the triple-phase boundary (TPB) and the surface of perovskite phases, thereby improving the oxygen permeability [44,47]. On the other hand, it may be concerned with the electronic conductivity of perovskite. Previous studies show that the $Pr_{0.6}Ca_{0.4}FeO_{3-\delta}$ (PCFO) exhibits the highest electronic conductivity among the Ca-based perovskite oxides [48], which is favorable for the oxygen permeability process according to the Wagner equation [49, 50]:

$$J_{O_2} = -\frac{RT}{16F^2L} \int_{\ln p_h}^{\ln p_l} \frac{\sigma_i \sigma_e}{\sigma_i + \sigma_e} d(\ln(p_o)) \qquad (6)$$

Where $J_{O2}$, $L$, $T$ are the oxygen permeation flux, membrane thickness and temperature, respectively. $R$, $F$ are the constant of gas and Faraday. $\sigma_i$ and $\sigma_e$ are the conductivity of electron and ion. $P_h$, $P_l$, $P_o$ signify the high oxygen partial pressure on the feed side, the low oxygen partial pressure on the sweep side, and the 1 bar standard oxygen partial pressure, respectively.

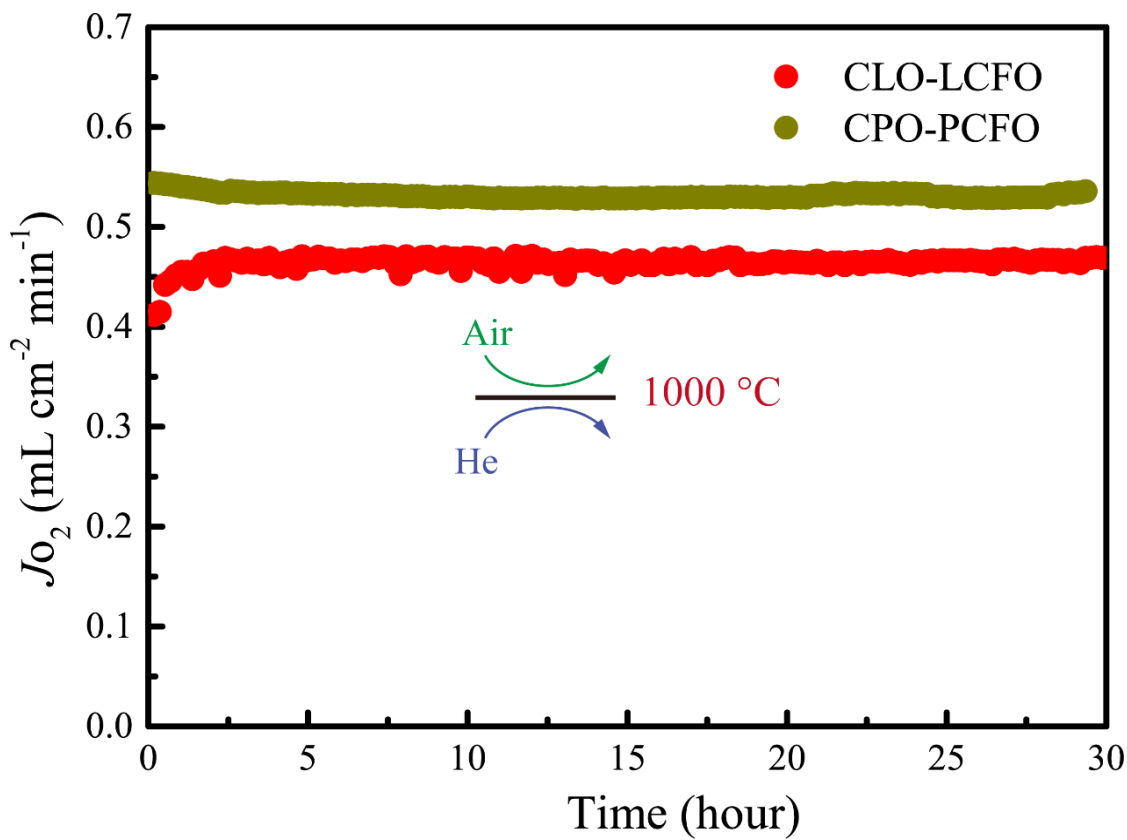

**Fig. 10** Oxygen permeation fluxes through 60wt.%$Ce_{0.9}Ln_{0.1}O_{2-\delta}$-40wt.%$Ln_{0.6}Ca_{0.4}FeO_{3-\delta}$ ($Ln$ = La, Pr) as a function of time with pure He as the sweep gas. *Conditions: membrane thickness: 0.6 mm; temperature: 1000 °C; 150 mL min$^{-1}$ air as the feed gas; 49 mL min$^{-1}$ He as sweep gas; 1 mL min$^{-1}$ Ne for calibration.* (Color online only)

**3.5 Long-term oxygen permeation**

At last, the long-term stability of these investigated four dual-phase membranes was investigated. **Fig. 10** shows that the evolution of time dependence of oxygen permeation flux under air/He gradient through these CPO-PCFO and CLO-LCFO membranes with higher oxygen permeability among our four studied compounds. It can be seen that both CPO-PCFO and CLO-LCFO dual-phase membranes can steadily work over 30 hours under air/He gradient at 1000 °C. In order to further estimate the probability of our investigated membranes used in the oxy-fuel combustion, the time dependence of oxygen permeation flux through four investigated membranes under the

$CO_2$ environment has also been studied. **Fig. 11** illustrates the long-term oxygen permeability of our investigated membranes under the $CO_2$ environment at different temperatures, in which the data of CPO-PCFO was extracted from ref. 40 [40]. When the temperature below 1000 ℃, the CSO-SCFO and CLO-LCFO membranes can stably work for 25 hours, whereas the oxygen permeation flux of CNO-NCFO membranes gradually reduces from 0.05 mL cm$^{-2}$ min$^{-1}$ to 0.035 mL cm$^{-2}$ min$^{-1}$. At 1000 °C, the oxygen permeation fluxes of CLO-LCFO and CNO-NCFO membranes sharply decline, while CSO-SCFO and CPO-PCFO membranes can be stably operated 25 hours under air/$CO_2$ gradient. To explore the specific reasons for the instability, the selected CNO-NCFO membranes after testing under Air/$CO_2$ gradient were further characterized. **Fig. S6** and **Fig. S7** are EDXS images of the feeding and sweeping sides of the spent CNO-NCFO membranes respectively. We can see that the elements are well-distributed in the feeding side, however, the aggregation of calcium element arises in the sweep side of the spent CNO-NCFO membrane. Similar phenomena have been reported in the previous literature [37]. Combined with the Ellingham diagram [27], in which the decomposition of carbonate is approximately 800 °C, it also implies the accumulation of calcium in the sweeping side attributes to the forming of calcium oxide rather than calcium carbonate. Therefore, we can conclude that the formation of calcium oxide in the spent CNO-NCFO under Air/$CO_2$ atmosphere maybe hinder the transportation of the oxygen ion, leading to the decease of the oxygen permeation flux.

Besides, the CPO-PCFO membrane exhibits much higher oxygen permeation flux than that of the CSO-SCFO membrane, which may be related to the fact that CPO has both higher ionic conductivity and non-negligible electronic conductivity [44]. As can be seen, the oxygen permeation flux through CPO-PCFO membrane can sustain around 0.25 mL cm$^{-2}$ min$^{-1}$ over the 40 hours, which is dissimilar with previously reported on the phenomenon of single-phase membranes, in which the oxygen permeation flux through $Ba_{0.5}Sr_{0.5}Fe_{0.2}Co_{0.8}O_{3-\delta}$ (BSFCO) membrane promptly drop from 1.5 mL cm$^{-2}$ min$^{-1}$ to 0 within 10 hours [21,25,51]. This result reveals that the CPO-PCFO membrane exhibits excellent $CO_2$ stability and comparable oxygen permeability, which is a vastly promising material candidate for the oxy-fuel combustion process.

Moreover, the average metal-oxygen bond energy (ABE) of the perovskite structure is an important physical parameter to evaluate $CO_2$ tolerance [52]. A higher ABE means that this membrane is more resistant to $CO_2$ corrosion [52,53]. Due to the lack of data in enthalpies of sublimation for Nd and La elements, we have calculated the ABE of perovskites in CPO-PCFO and CSO-SCFO membranes based on the data from thermodynamics handbook [33]. The ABE value (-318 kJ/mol) of PCFO is slightly higher than that of SCFO (-311 kJ/mol), implying the PCFO own better $CO_2$ stability. Similar, the ABE value (-318 kJ/mol) of PCFO is higher than that of BSFCO (-273 kJ/mol) [54]. This result is in good accordance with the aforementioned long-term stability test results, in which the oxygen permeation flux through the dual-phase membrane with the PCFO phase retains its original value over dozens of hours, whereas the oxygen permeation flux through the BSFCO membrane promptly drops immediately under pure $CO_2$ atmosphere.

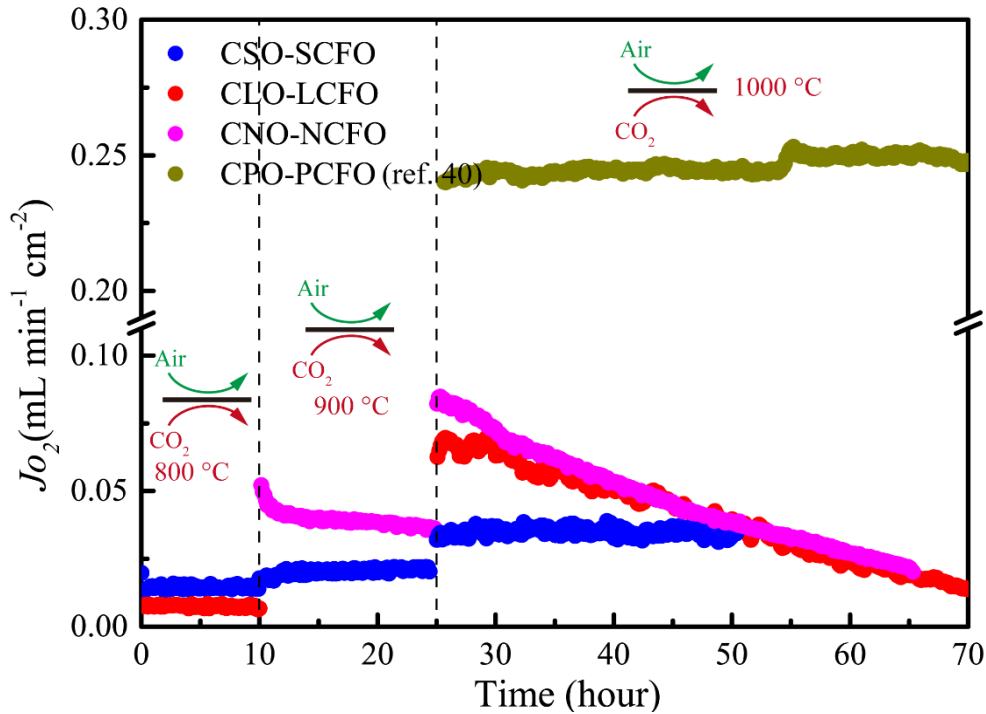

**Fig. 11** Oxygen permeation fluxes through $Ce_{0.9}Ln_{0.1}O_{2-\delta}$-40wt.%$Ln_{0.6}Ca_{0.4}FeO_{3-\delta}$ ($Ln$ = La, Pr, Nd, Sm) as a function of time with pure $CO_2$ as the sweep gas. *Conditions: membrane thickness: 0.6 mm; 150 mL min$^{-1}$ air as the feed gas; 49 mL min$^{-1}$ $CO_2$ as sweep gas; 1 mL min$^{-1}$ Ne for calibration.* (Color online only)

## 4. Conclusions

A series of 60wt.%$Ce_{0.9}Ln_{0.1}O_{2-\delta}$-40wt.%$Ln_{0.6}Ca_{0.4}FeO_{3-\delta}$ ($Ln$ = La, Pr, Nd, Sm) oxygen permeation composite membranes have been prepared successfully via a Pechini one-pot method. The influence of $Ln$ ($Ln$ = La, Pr, Nd, Sm) elements on the structure, oxygen permeability and stability are investigated systematically. Our results indicate that all membranes are composed of perovskite and fluorite two phases, however, there are two components in the perovskite phases of CNO-NCFO and CSO-SCFO membranes. Among all $Ce_{0.9}Ln_{0.1}O_{2-\delta}$-$Ln_{0.6}Ca_{0.4}FeO_{3-\delta}$ ($Ln$ = La, Pr, Nd, Sm) composite membranes, CPO-PCFO membrane displays the optimal oxygen permeation performance and excellent $CO_2$ tolerance. For CPO-PCFO, a steady oxygen flux of 0.25 mL cm$^{-2}$ min$^{-1}$ is sustained over 40 hours at 1000 °C under air/CO$_2$ gradient. This work provides some candidate OTMs for oxygen separation applied in the oxy-fuel combustion process and some ideas to search for new dual-phase membranes.

**Acknowledgment**

This work was supported by National Natural Science Foundation of China (No. 11922415 and No. 21701197), the Pearl River Scholarship Program of Guangdong Province Universities and Colleges (20191001)，the Fundamental Research Funds for the Central Universities (19lgzd03), Guangdong Basic and Applied Basic Research Foundation (2019A1515011718) and Key Research & Development Program of Guangdong Province (2019B110209003).

AUTHOR INFORMATIONS

Corresponding Authors

Tel.: (+0086)-2039386124, E-mail address: luohx7@mail.sysu.edu.cn (Huixia Luo)

Notes

The authors declare no competing financial interest.

**Highlights**

- New 60wt.%$Ce_{0.9}Ln_{0.1}O_{2-\delta}$-40wt.%$Ln_{0.6}Ca_{0.4}FeO_{3-\delta}$ dual-phase membranes are reported.

- These new Ca-containing membranes show comparable $CO_2$-tability and phase stability;

- Co-Free CLnO-LnCFO membranes are economic and ecological;

- CPO-PCFO shows highest oxygen permeability and stability among these compounds.

**Graphical abstract**

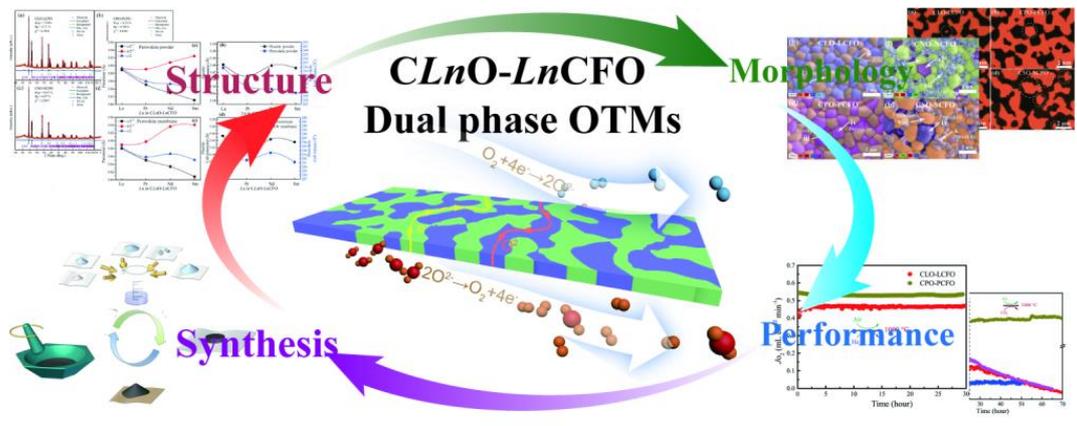

# Supporting information

# Influence of *Ln* elements (*Ln* = La, Pr, Nd, Sm) on the structure and oxygen permeability of Ca-containing dual-phase membranes

*Shu Wang[1], Lei Shi[1], Mebrouka Boubeche[1], Xiaopeng Wang[1], Lingyong Zeng[1], Haoqi Wang[1], Zhiang Xie[1], Wen Tan[1], Huixia Luo[1,2,3]\**

[1]*School of Materials Science and Engineering, Sun Yat-Sen University, No. 135, Xingang Xi Road, Guangzhou, 510275, P. R. China*

[2]*Key Laboratory for Polymeric Composite and Functional Materials of Ministry of Education, Sun Yat-Sen University, Guangzhou, 510275, China*

[3]*State Key Laboratory of Optoelecronic Materials and Technologies, Sun Yat-Sen University, Guangzhou, 510275, China*

*\*Corresponding author/authors complete details (Telephone; E-mail:)*
*(+0086)-2039386124*
*luohx7@mail.sysu.edu.cn*

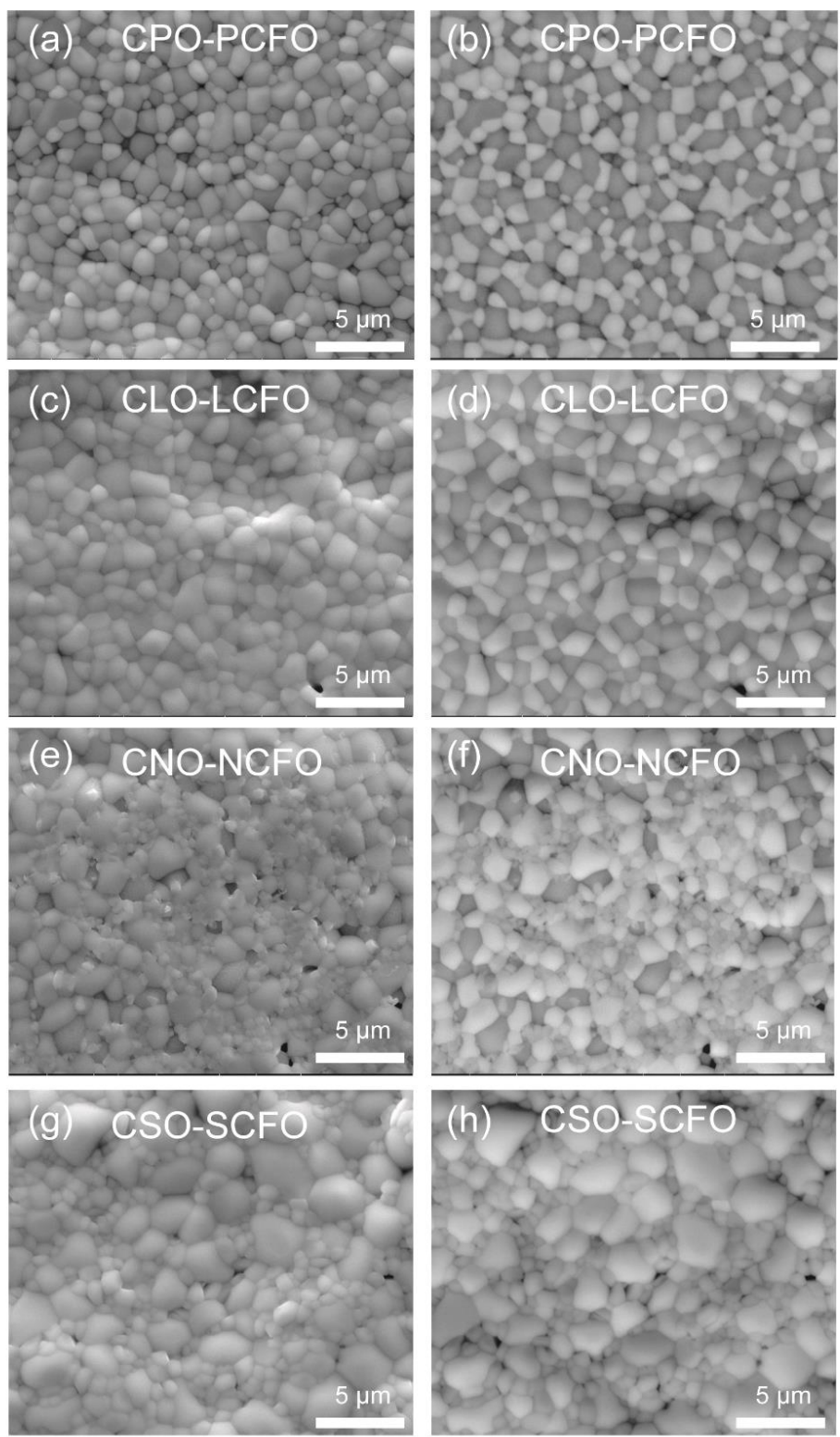

**Fig. S1** SE (a, c, e, g) and BSE (b, d, f, h) images of fresh 60wt.% $Ce_{0.9}Ln_{0.1}O_{2-\delta}$-40wt.%$Ln_{0.6}Ca_{0.4}FeO_{3-\delta}$ ($Ln$ = La, Pr, Nd, Sm) composites.

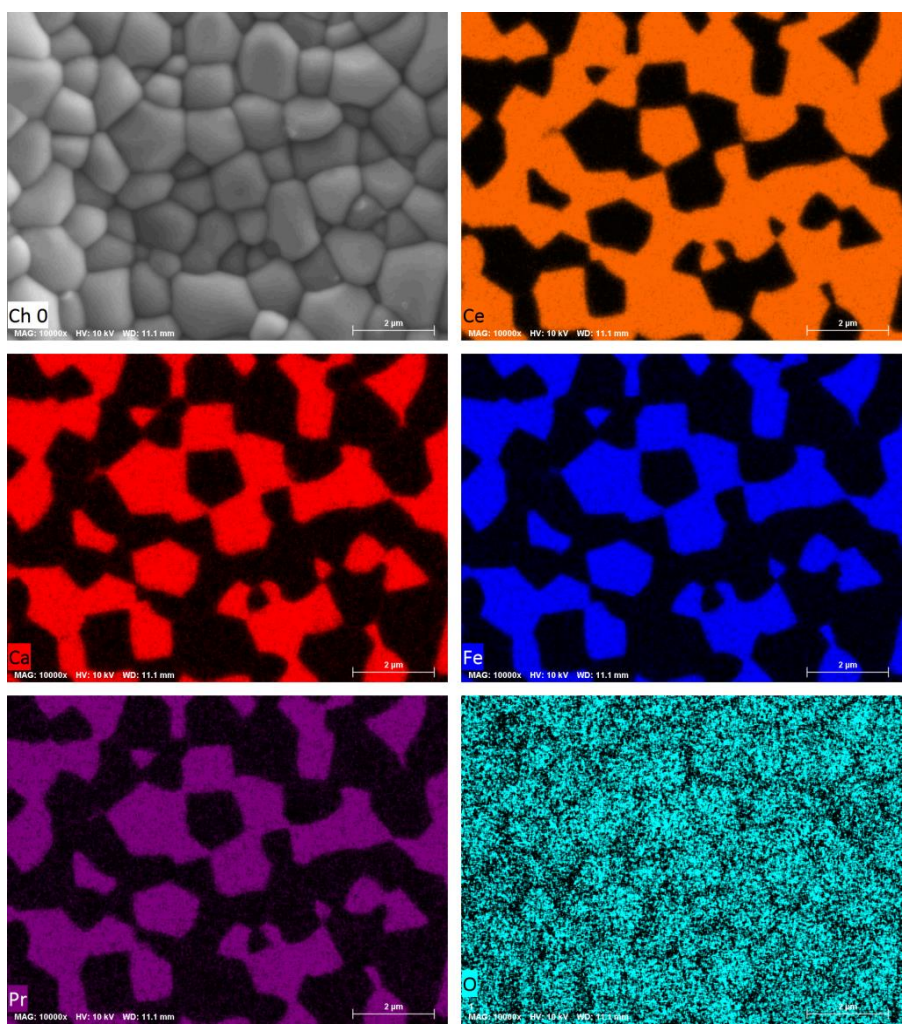

**Fig. S2** The EDXS mapping images of fresh CPO-PCFO membrane.

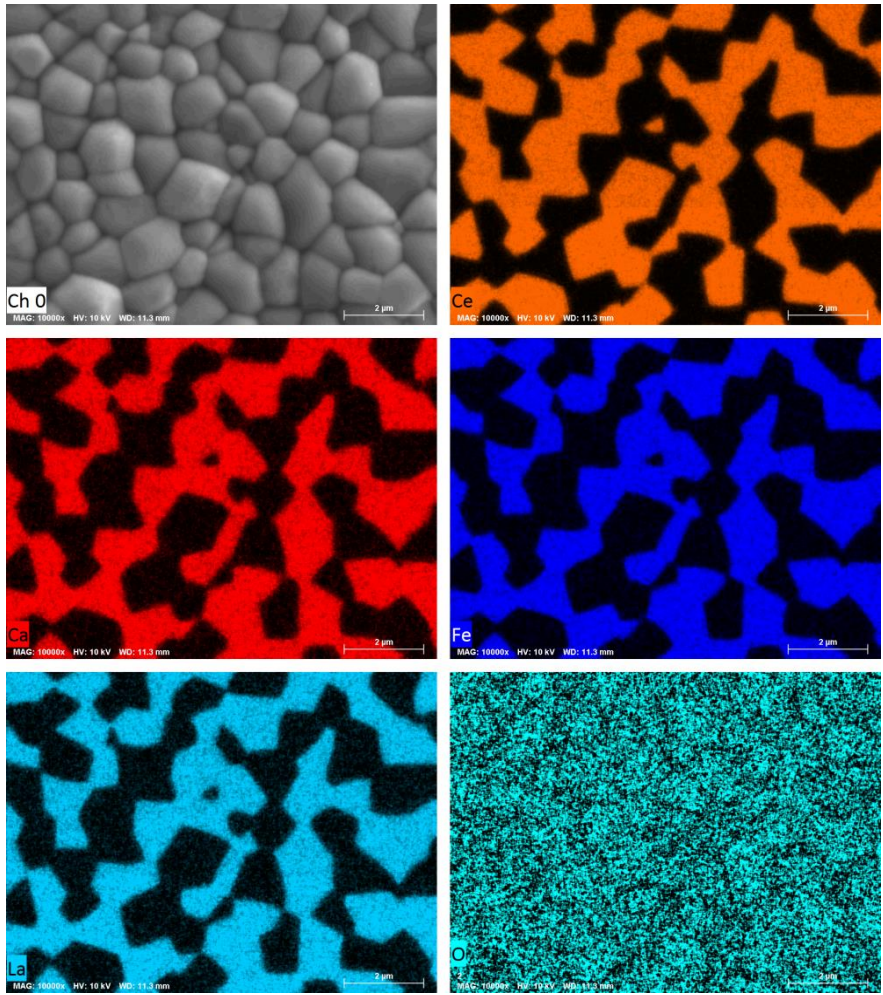

**Fig. S3** The EDXS mapping images of fresh CLO-LCFO membrane.

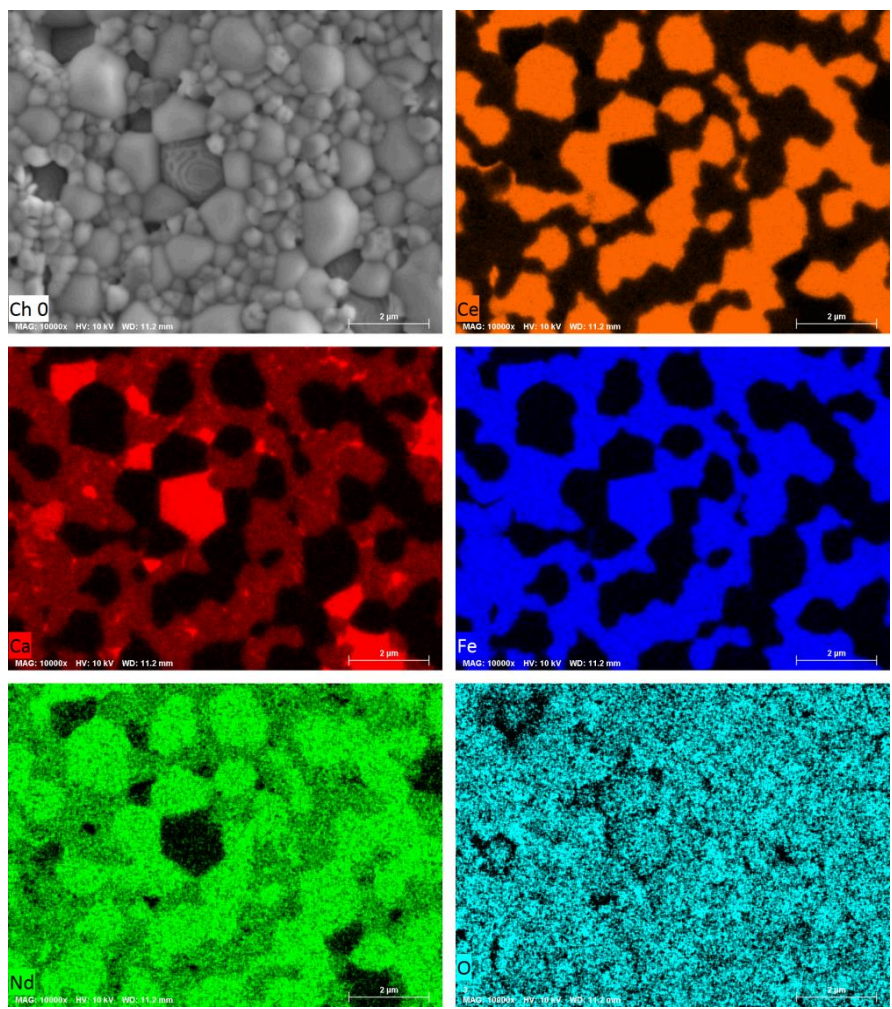

**Fig. S4** The EDXS mapping images of fresh CNO-NCFO membrane.

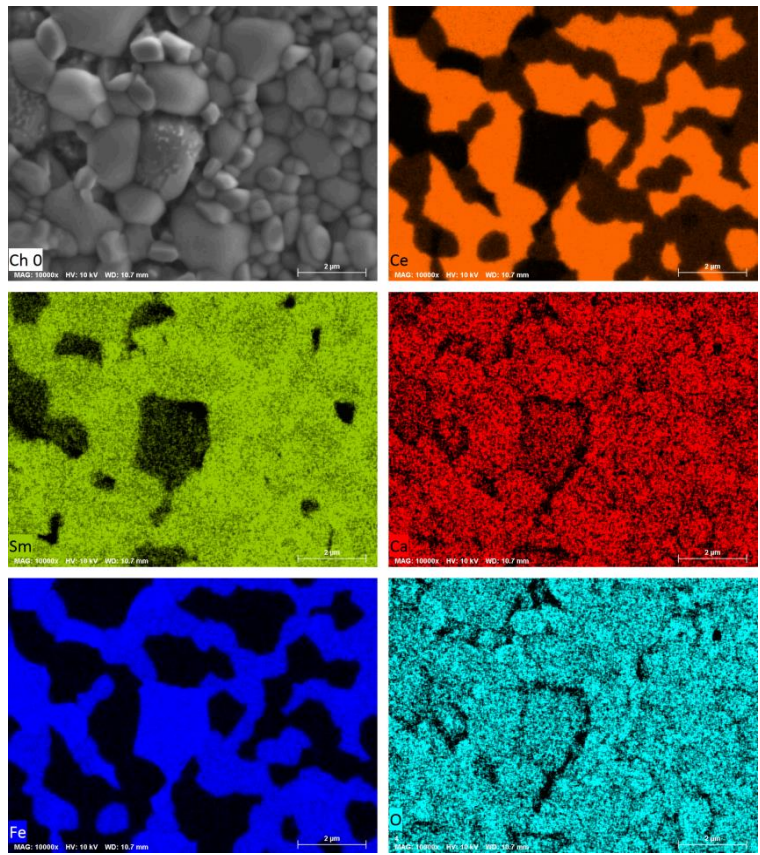

**Fig. S5** The EDXS mapping images of fresh CSO-SCFO membrane.

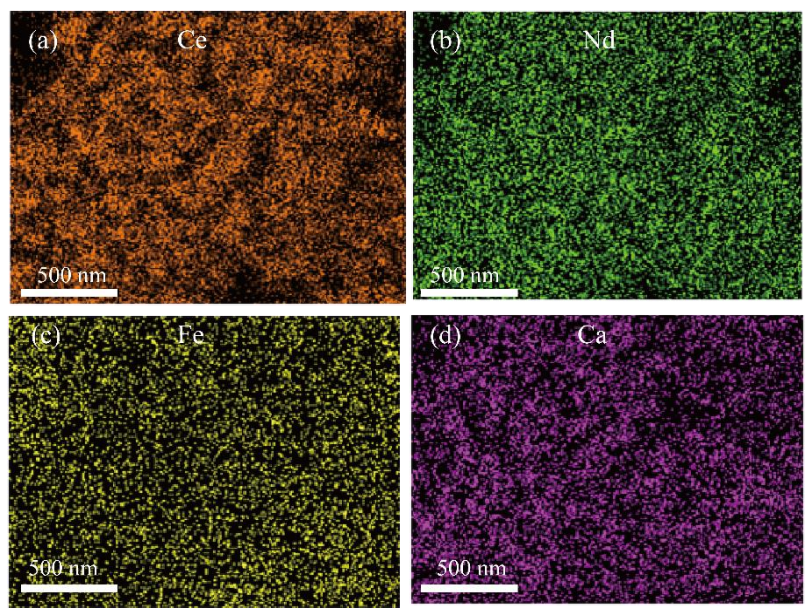

**Fig. S6** EDXS mappings for the feeding side of the spent CNO-NCFO membrane after long-term $CO_2$ stability tests.

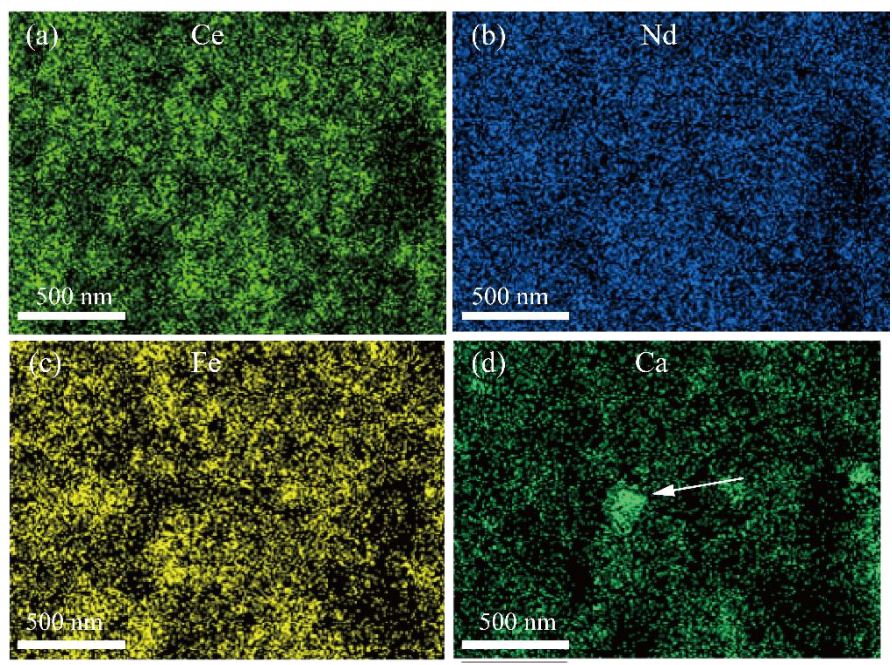

**Fig. S7** EDXS mappings with the sweeping side of the spent CNO-NCFO membrane after long-term $CO_2$ stability tests.